\documentclass[onecolumn,nofootinbib,aps]{revtex4}
\usepackage{graphicx}
\usepackage{dcolumn}
\usepackage{bm}
\usepackage{epsfig,epstopdf,amsmath}
\usepackage{amssymb}
\usepackage{amstext}
\usepackage{subfigure}
\usepackage{float}
\usepackage{multirow}
\usepackage{gensymb}
\usepackage{ulem}
\usepackage[usenames,dvipsnames]{color}
\usepackage[pagebackref=false, colorlinks=true]{hyperref}
\definecolor{redish}{rgb}{0.7,0.2,0.0}  
\definecolor{bluish}{rgb}{0.2,0.5,0.8}

\hypersetup{linkcolor=redish,          
	citecolor=blue,        
	filecolor=magenta,      
	urlcolor=bluish}          

\DeclareFontFamily{U}{rsfs}{}         
\DeclareFontShape{U}{rsfs}{m}{n}{<5> rsfs5 <6><7> rsfs7          %
	<8><9><10><10.95><12><14.4><17.28><20.74><24.88> rsfs10}{}     %
\DeclareMathAlphabet{\mathfs}{U}{rsfs}{m}{n}

\def \f{\frac}

\def \p{\partial}

\def \th{\theta}

\begin{document}
	\title{Accretion inside astrophysical objects : Effects of rotation and viscosity}	
	\author{H. A. Adarsha}
	\email{adarsha.mcnsmpl2023@learner.manipal.edu}
\affiliation{Manipal Centre for Natural Sciences, Manipal Academy of Higher Education, Manipal 576104, India}

  \author{Chandrachur Chakraborty}
\email{chandrachur.c@manipal.edu}
\affiliation{Manipal Centre for Natural Sciences, Manipal Academy of Higher Education, Manipal 576104, India}

\author{Sudip Bhattacharyya}
\email{sudip@tifr.res.in}
\affiliation{Department of Astronomy and Astrophysics,
Tata Institute of Fundamental Research, Mumbai 400005, India}

\begin{abstract}
Subsolar mass black holes could show up in gravitational wave observations in future and near-solar mass black holes might have been involved in the events GW190425 and GW190814.
Since they cannot form from the stellar evolution, their creation requires exotic mechanisms. 
One such mechanism involves the capture of dark matter particles by stellar objects and their thermalization.
When the criterion for 
the collapse of these dark matter particles
is satisfied, a tiny endoparasitic black hole (EBH) forms and then it accretes matter from the host.
The EBH may transmute the host into a black hole of nearly the same mass as the host or lesser, depending on the type of accretion.
We examine this complex and poorly-explored accretion mechanism, considering the effects of rotation and viscosity but ignoring some other effects, such as those of pressure and magnetic field, as the first step.
Using a general framework to assess the effects of rotation and viscosity on accretion, we show that the accretion could be stalled in some white dwarfs, but not in neutron stars.
The stalled accretion should cause an opening in the host's polar regions, the extent of which depends on the mass and spin of the host.
\end{abstract}	

\maketitle

\section{Introduction}
Stellar-mass black holes (BHs) generally form when the core of stars with masses exceeding 25$M_{\odot}$ \cite{Heger_2003} undergoes gravitational collapse.
However, near- and subsolar mass BHs cannot form by such a process. 
Nevertheless, gravitational wave observations have indicated the existence of such relatively low mass BHs  \cite{abbott_gw190425_2020,LVKcollab2023}. 
Thus, their origin requires an explanation. 
While these BHs could be primordial black holes (PBHs) \cite{miller2024gravitationalwavessubsolarmass}, the capture of a PBH by a stellar object, where the captured PBH accretes matter from the host and transmutes it into a BH, is another explanation \cite{GenoliniSerpicoTinyakov2021, baumgarte_primordial_2024}. 
However, the probability of such a capture is low \cite{Abramowicz_2009}.
On the other hand, plausible capture of dark matter (DM) particles by neutron stars, which thermalize and form a tiny endoparasitic black hole (EBH) upon collapse, is another idea that has been proposed in \cite{goldman_weakly_1989}.
Different calculations of the capture rate of DM have been done considering single scattering capture \cite{mcdermott_constraints_2012, kouvaris_constraining_2011}, multiscattering capture \cite{bramante_multiscatter_2017}, etc.
Such an EBH may accrete matter from the host and transmute it into a near- or subsolar mass BH or naked singularity \cite{Chakraborty_2024, Chakraborty_low_mass_nakedsingularity2024}. 

The end state of the process involving the capture of DM by stellar objects has been shown to be the transmutation of the host into a near-solar (or subsolar) mass BH which have nearly the same mass as their hosts (neutron stars or white dwarfs) and hence have similar distribution in mass range \cite{goldman_weakly_1989,gould_neuton_1990,kouvaris_constraining_2011,kouvaris_growth_2014,kouvaris_nonprimordial_2018,dasgupta_low_2021,baumgarte_max_surv_2021,takhistov_test_2021}. There are papers connecting the transmutation process with various observable signatures such as fast radio bursts, missing pulsar problem \cite{fuller2015}, quiet kilonovae \cite{bramante_searching_2018}, mass distribution of BHs identical to that of pulsar mass distribution in galactic bulges \cite{bramante_detecting_2014}, and gravitational waves \cite{east_fate_2019}. 
It has also been suggested that the transmutation process may create Kerr superspinars as end products in case of the transmutation of a white dwarf \cite{Chakraborty_2024}. 
These conclusions, however, are mostly drawn assuming a spherical Bondi accretion once an EBH forms.

However, a simple Bondi accretion (see \cite{kouvaris_constraining_2011} for example) of the host star's matter by the EBH may not be sufficiently realistic, since this does not take the spin of the host into account.
Most stellar objects have nonzero spin and hence the matter cannot be falling into EBH via radial trajectories. 
It has been shown that in case of accretion by an EBH inside Sun, the rotation (and viscosity) of the Sun imposes bounds on the mass of EBH, below (above) which accretion cannot be spherically symmetric \cite{d_markovic_evolution_1995}. 
These bounds, obtained for the equatorial plane of the host (as well as the EBH), do not significantly change the course of spherical Bondi accretion inside neutron stars (NSs) as well unless very extreme spin or rotation frequency and temperature are considered \cite{kouvaris_growth_2014}. 

The effect of rotation and viscosity on the accretion process is explained by considering the equatorial plane in \cite{d_markovic_evolution_1995} and \cite{kouvaris_growth_2014}. 
This is insufficient since the accretion happens outside the equatorial plane too. 
Here, we calculate the mass bounds on EBH between which the accretion can be stalled due to rotation of the host object. Since the bounds depend on polar angle $\theta$, we obtain an angle at which the two mass bounds are equal. The accretion is stalled at angles greater than this angle and continues below this angle. Further, we calculate the final mass of EBH which also gives the value of final angle at which accretion is stalled. This gives a configuration that has an opening near polar regions of the rotating host containing EBH within it, particularly for white dwarfs (WDs). We also show that it is hardly possible to obtain such a similar scenario in case of a rapidly spinning neutron star with spin rate $1200$ Hz which has not been observed so far.

This paper is organized as follows. In Sec. \ref{s2}, we discuss the general case of a rotating host body with an EBH and develop a general framework to find the effect of rotation and viscosity on the accretion process. Furthermore, we calculate the time required to attain the stalled accretion state. In Sec.  \ref{s3} we apply the formulation on neutron stars and white dwarfs and determine the opening angle, final mass of the EBH and the timescale of the stalling in white dwarfs and conclude in Sec. \ref{s4}.

\section{\label{s2}General formalism of accretion}

Here, we discuss the dark matter collapse, formation of EBH, and the accretion of the host's matter by the EBH considering the host's rotation and viscosity. 
However, we ignore the effect of pressure of host's matter \cite{d_markovic_evolution_1995, kouvaris_growth_2014} in the accretion process in this paper, which we will consider in a subsequent work. 

\subsection{Dark matter collapse and formation of endoparasitic black hole}
Dark matter (DM) particles can be captured by various stellar objects. 
This is expected to happen predominantly in regions where dark matter  density is very high, like the Galactic bulge. 
The capture rate $F$ \cite{mcdermott_constraints_2012, bramante_multiscatter_2017} for DM particles by an astrophysical object  is linearly dependent on DM density ($\rho_\chi$), the DM-nucleon scattering cross section ($\sigma_\chi$) and inversely proportional to velocity dispersion of DM particles ($\Bar{v}\approx$ 220 km/s) and mass of DM particle ($m_\chi$). 
One can consider different DM density profile models such as Navarro-Frenk-White (NFW) \cite{Navarro_1997}, Einasto \cite{Fabio_Iocco_2011} or cored DM density profile \cite{Olling_2001} to get the $\rho_\chi$ in capture rate.
Once enough DM is captured and the captured DM thermalizes with the host's matter, it forms a dark core at the center of the host. The DM core collapses to form an EBH, once the number of DM particles satisfies the condition $N\ge N_{\rm Ch}$ for fermionic DM and $N\ge N_{\rm self}$ for bosonic DM \cite{mcdermott_constraints_2012}, 
where $N, N_{\rm self}$, and $N_{\rm Ch}$ are number of DM particles, number of DM bosonic particles necessary for the onset of self-gravitation, and number of fermionic DM particles corresponding to the Chandrasekhar limit, respectively (see Sec. 2 of \cite{Chakraborty_low_mass_nakedsingularity2024} for details). 
The mass of the EBH depends on the mass of the DM particles as well as whether they are bosonic or fermionic. The formed EBH can have a minimum mass of $10^{15}$ g and can have up to $10^{31}$ g (corresponding to the mass of DM particles ranging from $10$ GeV to $10^6$ GeV) \cite{Chakraborty_2024} at its inception. 

An EBH thus formed starts growing by accreting mass from its host. The accretion process can at this stage be described to a good approximation by spherical Bondi accretion. 
However, at some stage of growth, angular momentum of infalling matter inevitably plays a role of slowing down the rate of accretion. 
The slowdown may lead to a complete stalling at specific instances as we show in this paper.

\subsection{\label{s2b}Conditions for stalling of accretion}
We consider the stalling of accretion by EBH owing to the excessive angular momentum of the infalling matter due to the rotation of the host. For the EBH to continuously accrete matter until the host is transmuted into a black hole (or a Kerr naked singularity \cite{Chakraborty_2024}), the infalling matter must have specific angular momentum lesser than the Keplerian specific angular momentum of innermost stable circular orbit (ISCO) of the EBH. 
Since all astrophysical objects should realistically have nonzero spin values, the infalling matter should definitely have angular momentum. 
If this angular momentum is not lesser than the angular momentum needed to sustain a stable circular orbit at the ISCO, then there should be some mechanism that transports the angular momentum of matter for it to fall into the EBH. 
Here, we consider viscosity as that mechanism. 
For simplicity, we also assume that the host is a rigidly rotating body.

Following \cite{d_markovic_evolution_1995} and \cite{ kouvaris_growth_2014}, we make a comparison between the specific angular momentum of a chunk of infalling matter and the specific angular momentum at ISCO of the EBH. 
This gives an upper bound for the mass of EBH, below which stalling due to excessive angular momentum of infalling matter occurs in the absence of a mechanism to transport angular momentum to outer regions.
This upper bound however is different for different polar angles ($\theta$), since for a rigidly rotating body, the specific angular momentum of a piece of matter at a higher latitude is lesser than that of a piece of matter in the equatorial plane. 
The spin of EBH increases in the course of its growth, but since the increase is initially slow, we consider Kerr parameter of EBH to be small ($a\rightarrow0$) but nonzero for now.
Hence, the necessary condition for stalled accretion is 
\begin{equation} 
    \ell>\ell_{\rm gISCO} \label{1}
\end{equation}
where $\ell$ is the specific angular momentum of a chunk of infalling matter and $\ell_{\rm gISCO}$ is the specific angular momentum at the innermost stable circular orbit of radius $r_{\rm gISCO}$ located in an arbitrary angle ($\theta$), i.e., in the off-equatorial plane of the said EBH. Such orbits on off-equatorial planes require adequate support from external forces and for this case, the pressure gradient components play this role. We refer to Appendix \ref{appendix} for a detailed discussion related to this issue. However, one can express the Keplerian specific angular momentum at $r_{\rm gISCO}$ as \footnote{In this paper, all the mathematical expressions are presented in the geometrized unit, setting Newton's gravitational constant ($G$) and velocity of light in vacuum ($c$) equal to one: $G=c=1$.}
\begin{equation}\label{2}
    \ell_{\rm gISCO} \approx 6M\, \sin\theta
\end{equation}
(see Appendix \ref{appendix} for the detailed derivation) where, $M$ is the mass of the black hole. The $\sin\theta$ term gives us the the Keplerian specific angular momentum at $r_{\rm gISCO}$ for all polar angles. In general, the specific angular momentum of a chunk of matter at a distance $r_o$ from the center of the host star, rotating with an angular velocity $\omega_o$ at a polar angle $\theta$ is given by $\ell=r_{\rm o}^2\omega_{\rm o}\,{\rm sin^2}\;\theta$. Such a chunk of matter can fall into EBH if its specific angular momentum is lesser than the Keplerian specific angular momentum, or else it forms stable orbits around EBH. 
Since the EBH would have grown to approximately the mass $M=4\pi\,r_{\rm o}^3\rho/3$ \cite{d_markovic_evolution_1995}, by the time a chunk of the host's matter reaches the innermost stable orbit of EBH,  $r_{\rm o}$ can be rewritten in terms of the density ($\rho$) of the host, i.e. $r_{\rm o}=(3\,M/(4\pi\rho))^{1/3}$. Using the condition in Eq. \eqref{1} and equating the right-hand side of Eq. \eqref{2} with $\ell=(3\,M/(4\pi\rho))^{2/3}\omega_{\rm o}\,{\rm sin^2}\;\theta$ we get the mass of EBH as:
\begin{equation}
    M_{\rm d}= \left(\frac{3}{4\pi \rho}\right)^2\left(\frac{\omega_{\rm o}\;\sin\theta}{6}\right)^3 \label{M d}.
\end{equation}
This acts as the upper bound on the mass of EBH, above which the angular momentum would not be able to stall the accretion.

So far, we did not take viscosity of the infalling matter into account. 
Viscosity can transport the angular momentum of infalling matter of the host star, thereby allowing accretion to proceed without stalling. 
It was shown in \cite{d_markovic_evolution_1995} that viscous braking imposes a roughly rigid rotation (angular velocity $\omega \approx \omega_o$) of fluids beyond the radius $r_\nu$ in stars, where
\begin{equation}\label{4}
    r_\nu= \frac{M^2}{c_{\rm s}^3 \nu}
\end{equation}
and that the angular velocity below this radius goes as $\omega=\omega_{\rm o}(r_\nu^2/r^2)$. Here $c_{\rm s}$ is the speed of sound in the host body's matter and $\nu$ is the kinematic viscosity of the host body's matter. When $r_\nu$ is lesser than the Bondi radius $R_{\rm S}=M/4c_{\rm s}^2$, the condition \eqref{1} is not satisfied inside any stellar objects. However, when $r_\nu$ is greater than the Bondi radius $R_{\rm S}$, we can say that there may arise a situation where the condition \eqref{1} is satisfied. Such a condition arises since the specific angular momentum of matter at any radius $r<r_\nu$ is $\ell=r_\nu^2\,\omega_{\rm o}\,\text{sin}\,\theta$. Therefore, we can find the mass of EBH corresponding to which $\ell=\ell_{\text{gISCO}}$. Equating the right-hand side of Eq. \eqref{2} with  $r_\nu^2\,\omega_{\rm o}\,\text{sin}\,\theta$, we get a lower bound on the mass of EBH, below which stalling of accretion is not possible:
\begin{equation}
M_\nu=c_{\rm s}^2\left(\frac{6\,\nu^2}{\omega_{\rm o}\; \sin\theta}\right)^{1/3} .\label{M nu}
\end{equation}
Now we have two limits on the mass of EBH, namely $M_{\rm d}$ of Eq. \eqref{M d} and $M_\nu$ of Eq. \eqref{M nu} between which accretion can come to a halt. Above $M_{\rm d}$ of Eq. \eqref{M d} angular momentum of matter is not enough to halt accretion, and below $M_\nu$ of Eq. \eqref{M nu} viscosity creates the condition where specific angular momentum of matter falling in is lower than $\ell_{\text{gISCO}}$. Now, the criterion for stalled accretion at any angle $\theta$ is given by:
\begin{equation}
    M_{\rm d}>M_\nu \label{6}.
\end{equation}
Note that the limiting masses $M_{\rm d}$ and $M_\nu$ are dependent on polar angle, which means that the values of $M_{\rm d}$ and $M_\nu$ are different for different planes corresponding to different values of $\theta$. This would mean that there are different mass ranges for different angles in which accretion is stalled. 

The varying mass limits from Eq. \eqref{M d} and Eq. \eqref{M nu} for different polar angles immediately implies that there is a value of the angle $\theta$ at which these two limits are equal even if $M_{\rm d}>M_\nu$ near the equatorial plane. This critical angle, below which the condition \eqref{6} reverses and accretion is not stalled, creates a conical region around the polar regions of the host, where accretion is at Bondi rate and the matter is accreted by the EBH to create a vacuum. It creates an opening in the polar regions of the host body with a half-opening angle:
\begin{equation}
    \theta_{\rm o} =  
    \sin ^{-1}\left(\frac{7.13~ \pi ^{3/5}  \nu^{1/5} \rho ^{3/5} c_s^{3/5}}{\omega_{\rm o} }\right)
    \label{9}.
\end{equation}
Although accretion is stalled at all angles greater than $\theta_{\rm o}$ for the time being, it continues at angles lesser than it. As the mass is added to EBH, it  grows and this increases the opening angle since the specific angular momentum required to resist accretion is now greater. This increase in opening angle can still be limited if the initial half-opening angle $\theta_{\rm o}$ is small enough. This allows us to calculate the final mass of the EBH.
One can show that the final mass of EBH is given by 
\begin{equation}
    M_{\rm f}=M_{\nu}^*+\delta M \label{Mebh}
\end{equation}
where, $M_{\nu}^*\equiv M_{\nu}|_{\theta=\theta_{\rm o}}$, and
\begin{equation}
    \delta M=(4/3)\pi\, R_{\rm h}^3\,\rho \,[1-\cos\theta_{\rm f}\,] \label{deltaM}.
\end{equation}
Here, $\theta_{\rm f}$ is the final half-opening angle and $R_{\rm h}$ is the radius of the host. The final half-opening angle is obtained from $M_{\rm d}$ since, the increase in mass of EBH from $M_\nu^*$ goes through the range of $M_{\rm d}$ corresponding to angles greater than $\theta_{\rm o}$. The final half-opening angle is then obtained as
\begin{equation}\label{10}
    \theta_{\rm f}=\sin ^{-1}\left(\frac{384^{1/3} \pi ^{2/3} \rho ^{2/3} M_{\rm f}^{1/3}}{\omega_{\rm o} }\right) \equiv \;{\rm sin^{-1}} \left(k \,M_{\rm f}^{1/3}\right)
\end{equation}
where $k=(384^{1/3} \pi ^{2/3} \rho ^{2/3})/\;\omega_{\rm o} $. The relation in Eq. \eqref{deltaM} is approximate because the volume we considered also includes the volume enclosed by the ISCO of the EBH. If we subtract that we get:
\begin{equation}
    M_\nu^*+(M_{\rm h}-M_\nu^*)\left[1-\text{cos}\,(\theta_{\rm f})\right]-M_{\rm f}=0 \label{final mass}
\end{equation}
where $M_{\rm h}$ is the mass of the host body. Eq. \eqref{final mass} gives us a complete picture of stalled accretion. 
The final mass of the EBH is related to the final half-opening angle $\theta_{\rm f}$ and the mass of the host. 
If we examine more closely, the final mass of EBH is dictated by the final opening angle which in turn depends on the mass, viscosity and rotational frequency of the host. 
The terms $M_\nu^*$ and $\theta_{\rm f}$, which are, respectively the mass of the EBH when the initial half-opening angle $\theta_{\rm o}$ is attained and the final half-opening angle, depend on the density of the host (which is dictated by its mass, provided there is a well-defined mass-radius relation for the host), rotational frequency and viscosity as can be read from Eq. \eqref{M nu} and Eq. \eqref{10}.
The equation describes the final state of either stalling or complete transmutation of the host after the mass $M_\nu^*$ is attained by EBH. At this point, the stalling that started at $\theta=\pi/2$ (equator) has reached its minimum possible stalling angle, i.e., $\theta_{\rm o}$ and further increased to finally settle at final half-opening angle $\theta_{\rm f}$. The equation is also valid in case of complete transmutation of the host, in which case, $M_{\rm f}$ is equal to the mass of the host $M_{\rm h}$. For complete stalling of accretion, it is necessary to have $M_{\rm h}\ll k^{-3}$.
In such a case, we get an analytical solution for Eq. \eqref{final mass} as
\begin{equation}
    M_{\rm f}\approx \frac{k^6\,M_{\rm h}^3}{8} =\frac{1.8\times10^4~\ \pi ^4 \,\rho ^4\,M_{\rm h}^3}{\omega _{\rm o}^6}.
    \label{Mf approx}
\end{equation}

\begin{figure}
    \centering
    \includegraphics[scale=0.5]{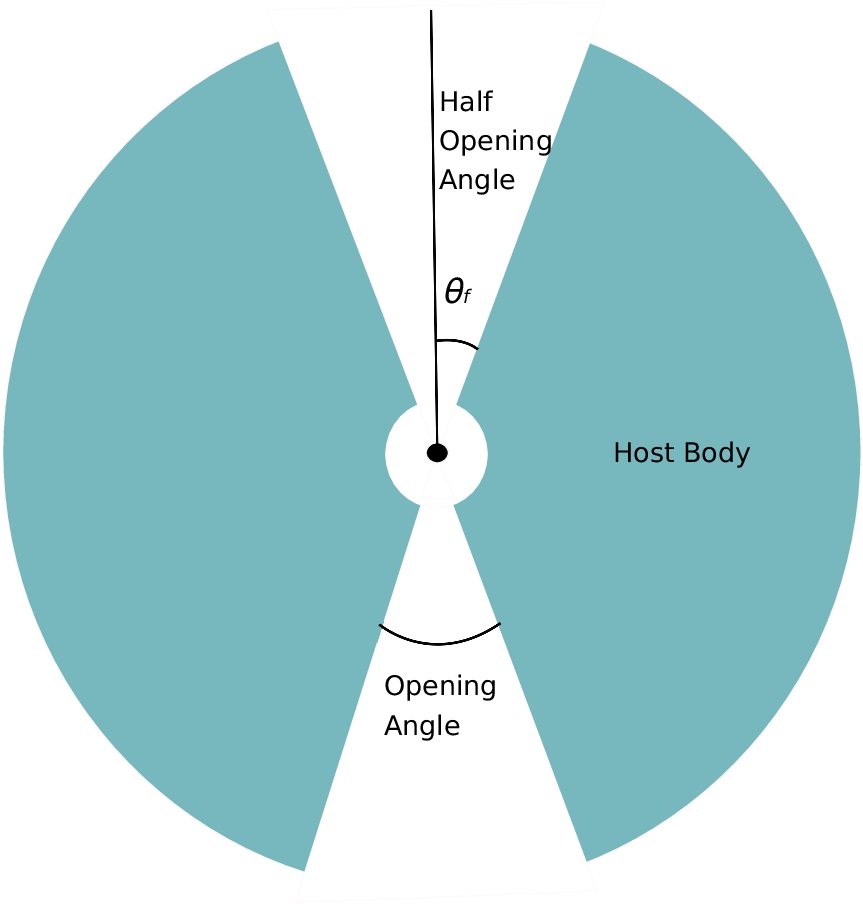}

\caption{The cartoon shows the basic idea of stalled accretion with conical openings of half-opening angle $\theta_{\rm f}$. The shaded part represents the host body and the black dot at the center depicts the EBH. For details, we refer to Sec. \ref{s2b}.}
    \label{fig:cartoon}
\end{figure}

Figure \ref{fig:cartoon} shows a cartoon of stalled accretion with an opening in a stellar object. In this figure, $\theta_{\rm f}$ is the half-opening angle which is obtained using Eq. \eqref{10}.

\subsection{Restart of accretion and an end to stalled accretion}
The stalled accretion condition could be sustained for long periods only if the mass of EBH by the end of its growth is not much greater than $M_\nu^*$. In other words, $M_{\rm h}\ll k^{-3}$ ensures completely stalled accretion. 
To get a clear picture the process, one can think of the accretion process from within the opening angle as a step-by-step process where, as the EBH accretes mass and grows, its $\ell_{\text{gISCO}}$ increases thereby increasing the angle $\theta$ from which it can accrete matter. 
If the mass accreted from within the initial opening angle $2\,\theta_{\rm o}$ is comparable to or greater than the mass of EBH at that time, it leads to a significant increase in the mass of EBH which in turn enables EBH to accrete matter with more angular momentum. This increases the opening angle further. The same holds for successive steps, i.e., if the addition to the mass of EBH is significantly large, the increase in $\ell_{\text{gISCO}}$ is also large, which leads to an increased opening angle. This process continues and it may eventually converge to a final opening angle that increases no further or may lead to complete transmutation depending on the mass, viscosity and rotation frequency of the host.

From Eq. \eqref{final mass}, one can deduce that for complete transmutation of the host, the condition $M_{\rm h}\rightarrow k^{-3}$ needs to be satisfied. When $M_{\rm h}< k^{-3}$, one can expect complete transmutation too unless the opening angle is significantly smaller than $\theta=\pi/2$.
Theoretically, the condition $M_{\rm h}\ll k^{-3}$ lays constraints on the rotation frequency, mass and viscosity of the host to get stalled accretion. Transmutation is inevitable if these conditions are not met.
Thus stalled accretion is expected in very tight conditions on rotation frequency, density, mass and viscosity of the host body. Only if the initial opening angle (Eq. \eqref{9}) is small enough that the mass that remains within it is lesser than $M_\nu^*$, the accretion remains stalled as the amount of mass added after reaching $M_\nu^*$ is very less and the final $M_{\rm EBH}$ is not significantly greater than $M_\nu^*$.

\subsection{\label{s2d}Timescales}
Astrophysical bodies like compact stars have a very long life time. Over this lifetime they can accrete particle DM and through the process of scattering and thermalization of DM with the baryonic matter in these bodies, DM accumulates at the core of these compact stars. Nonannihilating DM particles may have masses ranging from few keVs going up to $10^{12}$ GeV. Till about a mass of $10^{6}$ GeV, DM particles can be captured with a single scattering by the astrophysical body and then thermalizes with it. However, beyond this one needs to consider multiple scattering \cite{bramante_multiscatter_2017} followed by thermalization. For this work, we consider the single scattering capture of DM. The time taken for the whole process of capturing of DM and thermalization to self-gravitation and collapse amounts to formation timescale $t_{0}=N_{\rm Ch}/F\,+\,t_{\rm th}$ for fermionic DM and $t_{0}=N_{\rm self}/F\,+\,t_{\rm th}$ for bosonic DM \cite{Chakraborty_2024} where $F$ is the single scatter-capture rate of DM by astrophysical bodies.

The EBH starts to grow by accreting matter from the host as soon as it is formed.  
When the EBH has just formed, the accretion can be described by Bondi accretion as a good approximation since angular momentum of particles at the center of the host is much less than the $\ell_{\text{gISCO}}$ of EBH. Hence, till the EBH attains the mass $M_\nu^*$, at which point the lower and upper mass limits are equal ($M_{\rm d}=M_{\rm \nu}$) which corresponds to the initial opening angle $\theta_{\rm o}$, starting from initial mass $M_0$, one can take the rate of accretion to be equal to the Bondi accretion rate. The time it takes to do so is obtained by integrating the Bondi accretion rate formula: 
\begin{align}
    \frac{dM}{dt}= \frac{4\pi \rho}{c_{\rm s}^3}\,M^2\\
    \Rightarrow t_1=\int_{M_{0}}^{M_{\nu *}}\frac{c_{\rm s}^3}{4\pi \rho \,M^2}\\
    \Rightarrow t_1=\frac{c_{\rm s}^3\,(M_\nu^*-M_0)}{4\,\pi\,\rho\,M_0\,M_\nu^*} \label{timescale}
\end{align}
where $t_1$ is the accretion time till the stalling of accretion starts. After this point, the growth till it reaches $M_{\nu}^*$ is also through Bondi accretion but, reduced by a factor of $(1-\cos\theta_s)$, where $\theta_s$ is the angle where stalling has not happened yet. This growth is relatively fast and EBH soon attains the mass $M_{\nu}^*$ at which point, stalling is more or less complete for a fast rotating host, and any matter that remains within $\theta_{\rm o}$ will be accreted at Bondi rate by the EBH. This Bondi rate which now would be reduced to within the opening angle is
\begin{equation}
    \Dot{M}= \frac{4\pi \rho}{c_{\rm s}^3}\,M^2\,[1-\cos(\theta_{\rm o})]. \label{rate}
\end{equation}
The mass contained within the opening angle can be calculated using Eq. \eqref{deltaM} by replacing $\theta_{\rm f}$ with $\theta_{\rm o}$.
Even when the opening angle is very small, since the EBH is still a pointlike object in comparison to the size of the host, the volume of the conical opening is significantly large.
This means that the mass to be accreted from within this volume would be comparable, if not greater than the mass of EBH itself. With this in mind we can estimate the time it takes for EBH to accrete all the matter within the opening angle, to actually create the opening in the host's polar regions. The time it takes for the EBH to accrete all matter here is equal to the time it takes to grow from $M_{\nu}^*$ to the mass $M_{\rm f}$. Since $M_\nu$ varies as $1/(\sin\theta)^{1/3}$ (Eq. \eqref{M nu}) with polar angle, we can consider $M_\nu^*\approx M_\nu|_{\theta=\pi/2}\,(\equiv M_{\nu o})$. From Eq. \eqref{rate} we can write
\begin{equation}
    t_2\approx\frac{c_{\rm s}^3}{4\pi\rho\,\left[1-\cos(\theta_{\rm f})\right]}\int_{M_{\nu o}}^{M_{\rm f}} \frac{1}{M^2}\, dM \label{time}
\end{equation}
where $t_2$ is the time it takes for the EBH to accrete matter starting from the time accretion begins to be stalled on the equatorial plane, till the host settles in a stalled accretion state with conical openings of half-opening angle $=\theta_{\rm f}$ on both of its poles. One can easily see that this would be a very quick accretion after which the host would be left with an opening around its polar regions, which runs right down to its core, exposing the EBH to the outside, for our formalism and assumptions and in case of hosts which have the condition $M_{\rm h}\ll k^{-3}$ satisfied. 

\section{\label{s3}Applications : Stalling of accretion inside compact stars}
In this section, we apply the equations and relations formulated in the previous section, to see if accretion by EBH can be stalled inside compact stars with various rotation frequencies. We are considering only compact stars (e.g., neutron stars and white dwarfs) since their lifetime is long enough to accumulate enough DM and form EBH (see \cite{Chakraborty_low_mass_nakedsingularity2024} for additional reasons why compact objects are ideal cases for study). But the formulation can be applied to various kinds of astrophysical objects, including main sequence stars, planets, dwarf stars etc. 

\subsection{\label{s3a}Neutron Stars}
\begin{figure}[t]
    \centering
    \subfigure[~$\theta=\pi/2$]{\label{Ns M vs w}\includegraphics[scale=0.65]{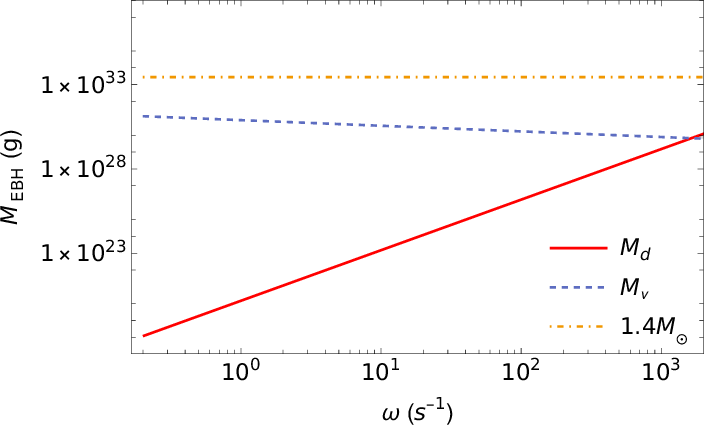}}\hspace{1.5cm}
    \subfigure[~$T_{\rm NS}=1.39$ ms \cite{Jason_Scott_716Hz}]
    {\label{Ns M vs ang} \includegraphics[scale=0.72]{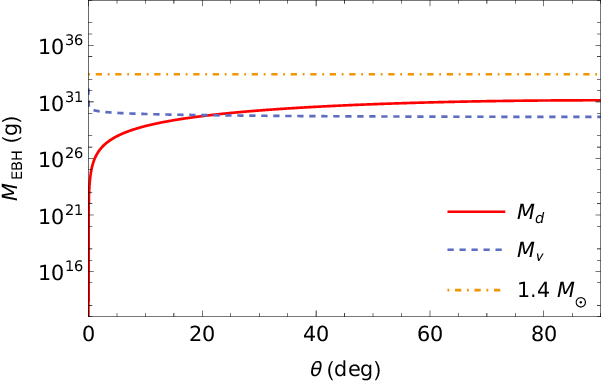}.}
    \caption{The limits on mass $M_{\rm d}$ and $M_\nu$ of EBH for stalled accretion plotted against angular frequency (panel (a)) and polar angle (panel (b)) for neutron star of $1.4\,M_{\odot}$.The red solid line represents the variation of upper mass limit $M_{\rm d}$ with angular frequency (panel (a)) or polar angles (panel (b)) and blue dashed line represents the variation of $M_\nu$ for the same. The orange dot-dash line shows the mass of the host which is $1.4\,M_{\odot}$. See Sec. \ref{s3a} for details.}
    \label{fig:NS M vs omega}
\end{figure}

Neutron stars have wide ranges of rotation frequencies and densities. The heaviest known neutron star has a mass of $\sim$ 2.35 \(M_\odot\) and a rotation period of 1.41 ms \cite{Romani_2022}, while the lightest one has a mass of $\sim$ 0.77 \(M_\odot\) \cite{doroshenko_strangely_2022} which is also a millisecond pulsar. The highest rotation period observed among radio pulsars is reported to be 53.8 minutes \cite{caleb_emission-state-switching_2024} while the lowest has a period of 1.39 ms \cite{Jason_Scott_716Hz}. These are the extreme values but we will consider typical values for density and rotation period i.e. a density of $4\times 10^{14}\;{\rm g\, cm^{-3}}$ and rotation period ranging from 1 ms to 10 second. The kinematic viscosity of neutron star is $2 \times 10^{11}\,{\rm cm^2\,s^{-1}}$ \cite{kouvaris_growth_2014}  for a temperature of $10^5$ K and the sound speed in neutron star is $\sim 0.17 c$ \cite{kouvaris_growth_2014}. In panel (a) of Fig. \ref{fig:NS M vs omega} we show the mass of EBH plotted against the angular velocity ($\omega \sim 1/T$ where T is the rotation period) for this range with the parameter values mentioned, for the equatorial plane (i.e. $\theta = \pi/2$) we see that the condition \eqref{9} is not satisfied for most of angular frequencies except at very high frequencies (barely visible in panel (a) of Fig. \ref{fig:NS M vs omega}). This can be interpreted to mean that viscosity efficiently evacuates the excess angular momentum of matter so that accretion proceeds almost spherically (i.e. Bondi accretion) even if EBH is very small and infalling matter has some excess angular momentum compared to $\ell_{\text{gISCO}}$. Hence we can safely say that a stalled accretion scenario is not to be expected in neutron stars. 

The mass limits $M_{\rm d}$ and $M_\nu$ plotted against the polar angle $\theta$ (panel (b) of Fig. \ref{fig:NS M vs omega}) also shows a similar picture. For a typical neutron star of density $4\times 10^{14}\;{\rm g\, cm^{-3}}$, rotation period of 1.39 ms and kinematic viscosity $2 \times 10^{11}\,{\rm cm^2\,s^{-1}}$, although it seems like there is a possibility of stalling, when we solve Eq. \eqref{Mebh} for this case, we find that it cannot sustain stalled state and gets transmuted with $M_{\rm f}=M_{\rm NS}$. This is due to the high density of neutron star which makes sure that the mass accreted from within the opening angle $\theta_{\rm o}$ is large enough to increase the mass of EBH to an extent where it can now accrete matter from even greater angles, close to equatorial plane. This ends up inevitably in complete transmutation of the neutron star into a black hole.

\subsection{\label{s3b}White Dwarfs}
\begin{figure}
    \centering
    \subfigure[$\:M_{\rm WD}=M_\odot$, $\theta=\pi/2$]{\label{WD M vs w}\includegraphics[scale=0.66]{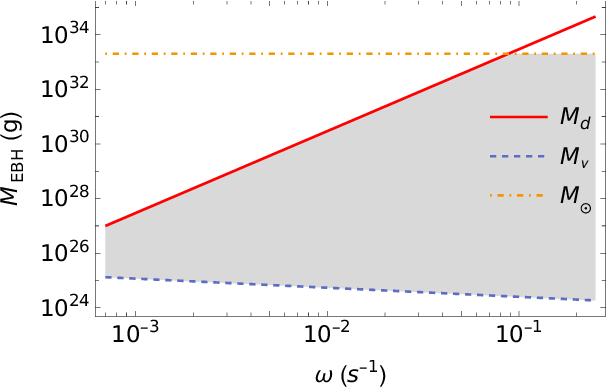}}\hspace{1.5cm}
    \subfigure[$\:M_{\rm WD}=0.17\,M_\odot$, $T_{\rm WD}=2000\,{\rm s}$]{\label{WD M vs angle 0.17M}\includegraphics[scale=0.68]{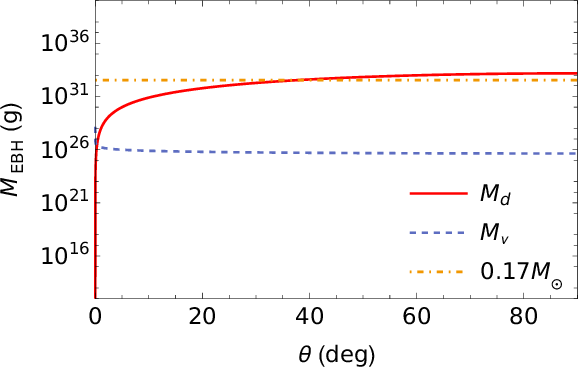}}\hspace{1.5cm}
    \subfigure[$\:M_{\rm WD}=0.6\,M_\odot$, $T_{\rm WD}=200\,{\rm s}$]{\label{WD M vs angle 0.6M}\includegraphics[scale=0.67]{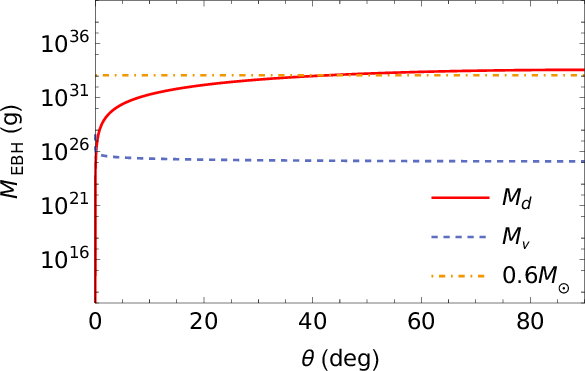}}\hspace{1.5cm}
    \subfigure[$\:M_{\rm WD}=1\,M_\odot$, $T_{\rm WD}=25\,{\rm s}$]{\label{WD M vs angle 1M}\includegraphics[scale=0.67]{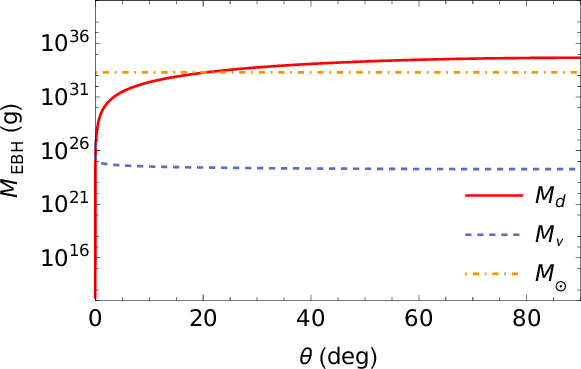}}\hspace{1.5cm}
    \subfigure[$\:M_{\rm WD}=1.3\,M_\odot$, $T_{\rm WD}=13\,{\rm s}$]{\label{WD M vs angle 1.3M}\includegraphics[scale=0.67]{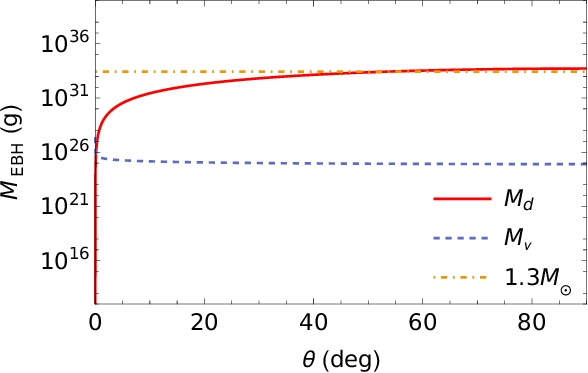}}\hspace{1.5cm}
    \caption{The limits on mass $M_{\rm d}$ and $M_\nu$ of EBH for stalled accretion plotted against angular frequency for the equatorial plane (panel (a)) and polar angle (panels (b)-(e)) for white dwarfs of different masses. The red solid line represents the variation of upper mass limit $M_{\rm d}$ with angular frequency (panel (a)) or polar angles (panels (b)-(e)) and blue dashed line represents the variation of $M_\nu$ for the same. The orange dot-dash line shows the mass of the host. For details, we refer Sec. \ref{s3b}.}
    \label{WD Md and Mv}
\end{figure}
White dwarfs are an interesting case, since it has been shown that if fast rotating white dwarfs undergo transmutation by EBH, they may form naked singularities with  Kerr parameter upto $a\approx19$ \cite{Chakraborty_2024}. Hence it is important to investigate whether such fast rotating white dwarfs can be fully transmuted by EBH accreting its matter at all. White dwarfs are known to have a wide range of rotation periods as well as a wide range of densities. The fastest rotating white dwarf observed so far has a rotation period of $\sim 25$ second \cite{25_sec_wd} whereas slowest ones may take hours to complete one rotation. The relation between density ($\rho_{\rm WD}$) of white dwarfs with its mass can be obtained from mass radius relation \cite{burrows_wd_massradius} as follows:
\begin{equation}
\rho_{\text{\tiny WD}} \approx \frac{(1.2\times 10^5) M_{\text{\tiny WD}}^2}{M_\odot R_\odot^3 \left[1-\left(\frac{M_{\text{\tiny WD}}}{M_{\text{Ch}}}\right)^{4/3}\right]^{3/2}} . \label{density}
\end{equation}
The kinematic viscosity of white dwarfs can be taken in the range $\nu\approx 10^2$ to $10^{10} \;{\rm cm^2\,s^{-1}}$ for white dwarfs of density in the range $10^4\,-\,10^{10}\;{\rm g\,cm^{-3}}$ 
(see \cite{dallosso_constraining_2014,Iben_Jr._1998} and the references therein) and the sound speed in white dwarf is $0.016c$ \cite{balberg_properties_2000}. Since the EBH has not accreted more than 1\% of the host's matter, the EBH would not have accreted a significant amount of host's angular momentum. Hence, the Kerr parameter of EBH still remains small ($a \rightarrow 0$) as considered in Sec. \ref{s2b}.
We consider uniform density across white dwarf obtained from Eq. \eqref{density} for white dwarf of mass $1M_\odot$ and plot the mass limits $M_{\rm d}$ and $M_\nu$ against angular velocity $\omega$ in panel (a) of Fig. \ref{WD Md and Mv} which shows us that accretion can be stalled for most of the spin periods. This plot however, only hints at the possibility as it plots for the equatorial plane. 

\begin{figure}[h]
    \subfigure[$\:M_{\rm WD}=0.6 M_\odot$, $T_{\rm WD}=140\,{\rm s}$, $\theta_{\rm f}\sim 3.1\degree$]{\label{WD M vs w}\includegraphics[scale=0.75]{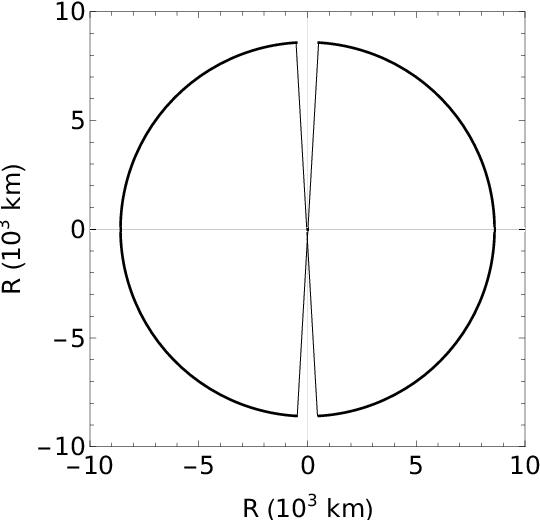}}\hspace{1.5cm}
    \subfigure[$\:M_{\rm WD}=0.6\,M_\odot$, $T_{\rm WD}=200\,{\rm s}$, $\theta_{\rm f}\sim9\degree$]{\label{WD M vs angle 0.17M}\includegraphics[scale=0.75]{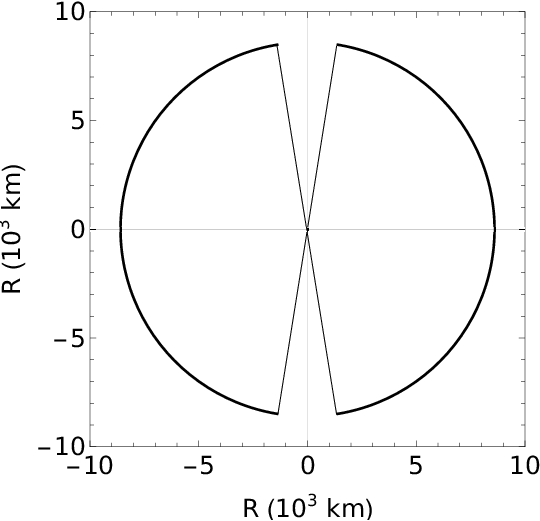}}
    
    \caption{The section of a white dwarf containing an endoparasitic black hole whose accretion has been stalled. The opening angle is small in (a) since $M_{\rm WD}=0.6 M_{\odot}$ and  $T_{\rm WD}\sim140$ s, whereas in (b),  $T_{\rm WD}\sim200$ s which is why the $\theta_{\rm f}$ is much larger compared to (a). For discussion refer Sec.  \ref{s3b}.}
    \label{fig:WD}
\end{figure}

In panels (b),(c),(d) and (e) of Fig. \ref{WD Md and Mv}, we see the mass limits $M_{\rm d}$ and $M_\nu$ plotted against polar angle $\theta$ for white dwarfs of mass $0.17M_\odot,\;0.6M_\odot,\;1M_\odot$ and $1.3M_\odot$ with time periods $T_{\rm WD}=$  2000, 200, 25 and 13 second in that order. The initial half-opening angle $\theta_{\rm o}$ corresponds to the meeting point of the curves of $M_{\rm d}$ and $M_\nu$ plotted against $\theta$. We can see that the opening angle is indeed very small for most of the chosen parameters. But to really see if such a situation is sustainable we need to solve Eq. \eqref{final mass} numerically and see whether stalling is sustainable. If the opening angle is very small, we can also use Eq. \eqref{Mf approx} to get the final mass of EBH when accretion has come to a complete halt.
\begin{center}
\begin{table}
\small
    \centering
    \begin{tabular}{ m{2.5cm}  m{2.5cm} m{2.5cm} m{2.5cm} m{2.5cm}} 
        \hline
        ${\rm M}_{\rm WD}\;(M_\odot)$&$\rho\;({\rm g\;cm}^{-3})$&${\rm T}_{\rm WD}$ (s)&$M_{\rm f}\;(M_\odot)$&$\theta_{\rm f}$ (deg)\\ \hline 
        \multirow{4}{*}{0.17 \cite{Kilic_2007}}& \multirow{4}{*}{$3.16\times10^{4}$}&50&$5.2\times10^{-12}$&0.0003\\ 
        {}&{}&500&$1.1\times10^{-6}$&0.20 \\
        {}&{}&2000&$4.6\times10^{-3}$&13 \\
        {}&{}&2500&$2.0\times10^{-2}$&27 \\
        \hline 
        \multirow{4}{*}{0.6}& \multirow{4}{*}{$4.41\times10^{5}$}&25 \cite{25_sec_wd}&$3.9\times10^{-8}$&0.02\\ 
        {}&{}&100&$1.2\times10^{-4}$&1.12 \\
        {}&{}&200&$7.5\times10^{-3}$&9.03 \\
        {}&{}&290&$8.6\times10^{-2}$&29 \\
        \hline 
        \multirow{4}{*}{1}& \multirow{4}{*}{$2.86\times10^{6}$}&13 \cite{Mereghetti}&$4.6\times10^{-6}$&0.17\\ 
        {}&{}&25&$2.3\times10^{-4}$&1.23 \\
        {}&{}&50&$1.5\times10^{-2}$&10 \\
        {}&{}&70&$1.3\times10^{-1}$&29 \\
        \hline 
        \multirow{4}{*}{1.3 \cite{caiazzo_highly_2021}}& \multirow{4}{*}{$2.27\times10^{7}$}&10&$8.3\times10^{-3}$&6.5 \\ 
        {}&{}&13\cite{Mereghetti}&$4.2\times10^{-2}$&14.4\\
        {}&{}&16&$1.6\times10^{-1}$&28 \\ 
        \hline
    \end{tabular}
    \caption{Final mass ($M_{\rm f}$) and final half-opening angle ($\theta_{\rm f}$) for some white dwarfs of masses ($M_{\rm WD}$) in the whole observed range and observed spin period ($T_{\rm WD}$). Some of these values are taken from real observed white dwarfs. 
    For details, we refer Sec.  \ref{s3b}.}
    \label{Table}
\end{table}
\end{center}
Table \ref{Table} contains some of the solutions for final mass and final half-opening angle for some of selected mass of white dwarfs. The table shows us that the final mass of the EBH in case accretion is stalled, is dependent on the spin period and the density of the white dwarf (which in turn depends on the mass of white dwarf). The mass of EBH depends on the viscosity too as $M_\nu^*$ depends on it. We also see an increasing opening half angle $\theta_{\rm f}/2$ as the time period of rotation of the host of a given mass increases. This is expected since the stalling is supported by excess angular momentum of host's matter. The opening angle is very small for fast rotating white dwarf but as the rotation period increases, the opening angle becomes larger at a very fast pace. Fig. \ref{fig:WD} shows a section view of how the stalled accretion would look with a half-opening angle of $\sim 3.1\degree$ for $M_{\rm WD}=0.6 M_{\odot}$ and $T_{\rm WD}\sim140$ s in panel (a), whereas in panel (b), $M_{\rm WD}=0.6 M_{\odot}$ and $T_{\rm WD}\sim200$ s which makes a larger opening angle of $\sim 9\degree$. 

\begin{figure}[h]
    \centering
    \includegraphics[scale=0.8]{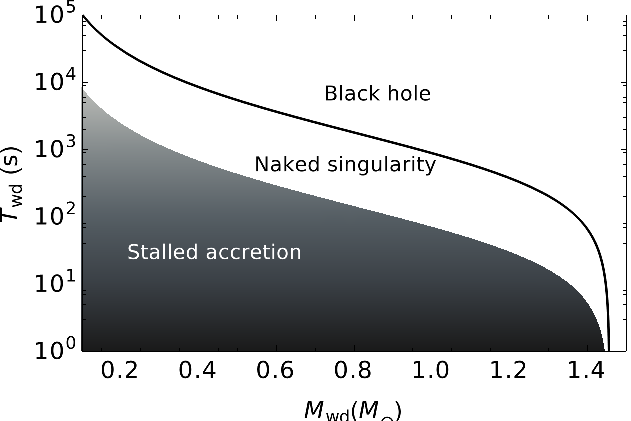}
    \caption{The mass vs time period space of white dwarfs. The end states of white dwarfs with EBH are shown labeled. The solid black line between the regions labeled as Black Hole and Naked Singularity, is the $a=1$ curve.
    The shaded region is where there is chance of stalling of accretion, the region below the $a=1$ curve and above shaded region white dwarfs transmute to naked singularities. Above the $a=1$ curve, white dwarfs tranmute into BHs. Refer Sec.  \ref{s3b} for details.}
    \label{fig:Domain of stalling}
\end{figure}

The possibility of stalling of accretion in white dwarfs takes away a significant portion of the possible naked singularity formation region as depicted in Fig. 1 of \cite{Chakraborty_2024}. We take the condition $M_{\rm WD}\lesssim \,k^{-3}$ and plot the condition in $T_{\rm WD}\;\text{vs}\;M_{\rm WD}$ space which gives Fig. \ref{fig:Domain of stalling}. In white dwarfs of high spin frequency, accretion is seen to be stalled in the whole range of mass as shown in Fig. \ref{fig:Domain of stalling}. This includes the portion that may or may not remain stalled, but gives us a region where naked singularity formation may potentially be prevented. The naked singularity formation may be prevented in cases where the transmuted object has a chance of forming a naked singularity with a large Kerr parameter ($a\sim19$) \cite{Chakraborty_2024}.

\begin{figure}[h]
    \centering
    \subfigure[$\;\rho_\chi^{\rm NFW}=8\times10^4\;\text{Gev cm}^{-3}$]{\label{WD M vs w}\includegraphics[scale=0.80]{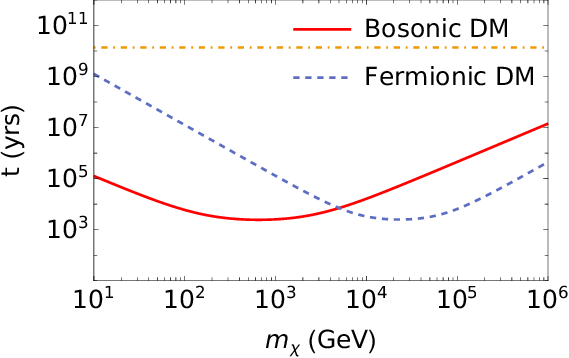}}\hspace{1.5cm}    \subfigure[$\;\rho_\chi^{\rm Einasto}=1\times10^4\;\text{Gev cm}^{-3}$]{\label{WD M vs angle 0.17M}\includegraphics[scale=0.80]{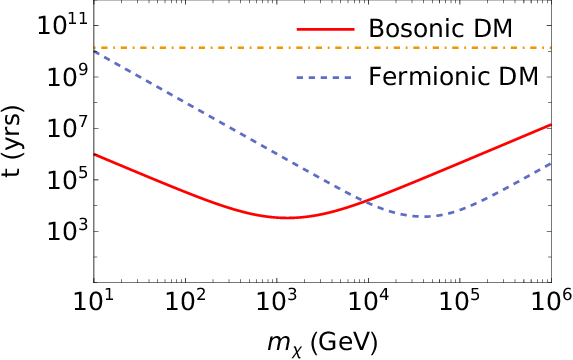}} \hspace{1.5cm}  \subfigure[$\;\rho_\chi^{\rm Cored}=17\;\text{Gev cm}^{-3}$]{\label{WD M vs angle 0.17M}\includegraphics[scale=0.80]{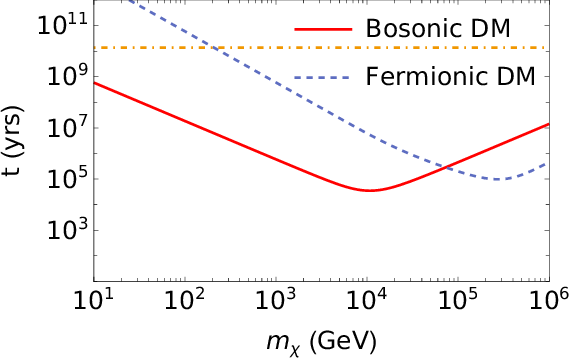}} \hspace{1.5cm}
    \subfigure[$\;\rho_\chi^{\rm Cored}=410\;\text{Gev cm}^{-3}$]{\label{WD M vs angle 0.17M}\includegraphics[scale=0.80]{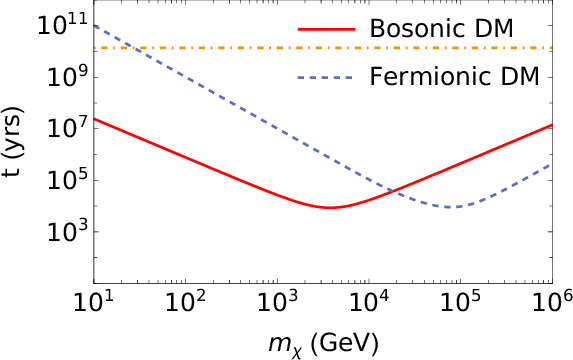}}
    \caption{Time required to get stalled accretion for the white dwarfs located in the Galactic centre with $M_{\rm WD}=0.6\;M_{\odot}$ and $T_{\rm WD}=25$ second. We assume the DM-Nucleon scattering cross section $\sigma=10^{-41}\;{\rm cm}^{-2}$ \cite{Chakraborty_2024}. The timescales differ for bosonic and fermionic DM (depicted by red solid line and blue dashed lines respectively). The orange dot-dash line shows the age of the universe. For detailed discussion we refer Sec. \ref{s3b}}
    \label{fig:stalling time}
\end{figure}

The time it takes for the white dwarf to accumulate DM and form an EBH can be estimated using $t_0=\text{Min}[N_{\rm Ch},N_{\rm self}]/F+t_{\rm th}$.
Once the EBH is formed, it starts accreting matter from the white dwarf at Bondi rate. The time to grow till $M_\nu|_{\theta=\pi/2}$ can be estimated using Eq. \eqref{timescale}.
In cases where opening angle is very small, the mass within that volume is also small, so that the opening angle does not increase further drastically. Using Eq. \eqref{time} we can estimate the time required for EBH to accrete matter from this region to create an actual opening around the polar regions of a rotating white dwarf. For example, Fig. \ref{fig:stalling time} shows the total time $(t \equiv t_0+t_1+t_2)$ till complete stalling of accretion starting from capture of DM, thermalization, formation of EBH and accretion from host for the DM particles (both boson and fermion) inside a white dwarf of mass $0.6\,M_{\odot}$ with temperature of $10^6$ K and spin period of 25 second, located in the Galactic center. We follow \cite{Chakraborty_2024} and estimate the timescale for various DM profiles like NFW \cite{Navarro_1997}, Einasto \cite{Fabio_Iocco_2011} and cored \cite{Olling_2001} DM profiles (see Eqs. 3.3--3.5 
of \cite{Chakraborty_2024} and related discussion there) for the same mass and time period of white dwarf. The timescale is well within the age ($t_U$) of Universe \cite{Planck2018} as well as within the expected age of many white dwarfs. 
The interesting fall and rise of the total timescale as a function of increasing DM mass is due to the varying formation timescale and accretion timescale. The formation time $t_0$ includes DM particles capture time and thermalization time, as discussed earlier. The DM particle capture time is inversely proportional to the DM mass since the Chandrasekhar limit (see Eq. (2.4) and Eq. (2.6) of \cite{Chakraborty_low_mass_nakedsingularity2024} and Eq. (3.2) of \cite{Chakraborty_2024}) which decides the number of DM particles necessary for collapse, is inversely proportional to the DM particle mass. This also means that, for heavier DM, the initial mass $M_0$ of formed EBH is smaller (see Fig. 1 of \cite{Chakraborty_low_mass_nakedsingularity2024}). 
Since accretion time $t_1+t_2$ is inversely proportional to the initial mass $M_0$ accretion timescale dominates for heavier DM, whereas $t_0$ dominates for lighter DM particle. 
Thermalization time on the other hand is longer for larger DM masses and smaller DM-baryon cross sections \cite{kouvaris_constraining_2011}. However when compared with capture time and accretion time, we can say that it is not very significant although it can slightly increase the formation time $t_0$ for larger DM masses (one may also look into Sec. 2.2 of \cite{Chakraborty_2024}). Thermalization time, therefore, tends to  increase the total timescale for the whole range of DM mass, albeit more for larger DM mass than smaller. 
The interplay between formation time, thermalization time and the accretion time gives rise to the trend in Fig. \ref{fig:stalling time}.

\section{\label{s4}Conclusion and Discussion}
In this paper, we have put forth the idea of stalled accretion and shown how such a state may be realized in the astrophysical context. The realization of this state in any rotating stellar object leads to the phenomenon of `conical opening' wherein the host undergoes a transformation into a structure with conical openings with an opening angle dictated by the mass, angular momentum and viscosity of the host, running down to the core along its poles. When applied to the compact stars, it is seen that the white dwarfs can indeed attain such a state given the assumptions. The same effect is not observed in neutron stars as the configuration cannot sustain itself, which was hinted at in \cite{kouvaris_growth_2014} where it is shown that neutron stars transmute despite their angular momentum. Now, if the stalled accretion really occurs in a white dwarf, this means that the particular white dwarf contains an EBH in its core but cannot be transmuted to a collapsed object (black hole or naked singularity).
This also indicates that there could be a possibility that not all the white dwarfs in the Galactic center can be transmuted to collapsed objects unlike the pulsars, even if the white dwarfs capture the DM particles and form EBHs inside them. Instead, some white dwarfs could appear as the regular white dwarfs for the distant observers despite having the presence of EBH in its core since they could sustain the stalled accretion scenario inside them.  This cannot arise in case of the neutron stars as shown in this paper.
It is important to note here that the total mass loss ($M_{\rm H}$) due to the Hawking radiation in the time $(t_U-t_0)$ is negligibly small (see Eq. A7 of \cite{C_Chakraborty_22_gravmag} with $n=0$) compared to the value of $M_{\rm f}$ (see Table \ref{Table}). For example, a EBH with $M_{\rm f} \sim 10^{-12}M_{\odot}$ ($10^{-2}M_{\odot}$) loses only $M_{\rm H} \sim 1$ g ($10^{-20}$g)
until now after its formation. Thus, the white dwarfs should be able to sustain the stalled accretion in the current Universe, as they still contain the EBHs in their core.

In this paper, the effect of pressure of host's matter in the accretion process is neglected \footnote{Ignoring the effect of pressure in the accretion process does not mean that there is no pressure support in the host NS/WD. The existence of NS/WD suggests that the gravitational pressure is balanced by the neutron/electron degeneracy pressure inside it, that leads to give a stable structure of such a compact object like NS/WD.  If there is no pressure support at all, such a stable compact object (NS/WD) would cease to exist. Thus, it indicates that the degeneracy pressure indirectly makes the circular obits possible to exist in the off-equatorial plane around the rotational axis with $\theta$=constant.} following \cite{d_markovic_evolution_1995, kouvaris_growth_2014}.
Qualitatively, one can say that for stalled accretion to be realistically possible, there needs to be a delicate balance between the degeneracy pressure of electrons/neutrons in the host WDs/NSs, the gravitational pressure and pressure due to the centrifugal forces. This is necessary on every surface that forms between the host and the EBH in the process of stalled accretion, as well as on the walls of conical opening. 
However, if such a balance of various pressures is not formed on the walls of conical opening, then matter from the equatorial region may flow towards the polar region. 
In such a case, the accretion
will not be stalled, but it will be a new and interesting kind of
accretion not explored so far. 
Such an accretion requires
an involved and detailed treatment, and hence we will consider it in a subsequent paper.
The accretion inside astrophysical objects is a complex, poorly explored, but extremely interesting
and important field, and, we plan to approach this problem systematically and 
step-by-step.
Hence, we consider two
physical effects, viz., rotation and viscosity, in this paper, and will
consider other effects, such as pressure and magnetic field, in subsequent
papers.
Consideration of these additional physical effects
in the future will give more realistic results, but our current
analysis will remain an indispensable component of any future
analysis, will be useful to gain insight, and will be very important
to interpret any future results.

\vspace{0.5cm}

{\bf Acknowledgements : }
HAA acknowledges Dr. TMA Pai Ph. D scholarship program of Manipal Academy of Higher Education.
SB acknowledges the support of the Department of Atomic Energy (DAE), India. We thank the referee for constructive comments that helped to improve the manuscript.
\\

\appendix

\section{\label{appendix}Derivation of the angular momentum at the innermost stable circular orbit located in the arbitrary plane of constant angle}~

Let us consider the line element for a general stationary axisymmetric spacetime, which can be expressed as
\begin{align}\label{A1}
ds^2=\textsl{g}_{tt}\,dt^2+2\textsl{g}_{t\phi}\,dt\,d\phi+\textsl{g}_{\phi\phi}\,d\phi^2+\textsl{g}_{rr}\,dr^2+\textsl{g}_{\theta\theta}\,d\theta^2 
\end{align}
where all the metric components of $\textsl{g}_{\mu \nu}$ are the functions of $(r,~ \theta)$. Now if a circular orbit is located in a particular angle $\theta$, equatorial plane or parallel to the equatorial plane, i.e., in the off-equatorial plane of the said spacetime due to the presence of a non-gravitational force, the last term of Eq. (\ref{A1}) vanishes due to $\theta =$ constant. The other metric components of $\textsl{g}_{\mu \nu}$ change accordingly for a constant value of $\theta$.
The standard normalization condition for the four-velocity ($\dot{x}^\mu \equiv [\dot{t}, \dot{\phi}, \dot{r}, \dot{\theta}]$) of a test particle  is expressed as \cite{Ryan1995}
\begin{eqnarray}
\textsl{g}_{tt}\,\dot{t}^2+2\textsl{g}_{t\phi}\,\dot{t}\,\dot{\phi}+\textsl{g}_{\phi\phi}\,\dot{\phi}^2+\textsl{g}_{rr}\,\dot{r}^2+\textsl{g}_{\theta\theta}\,\dot{\theta}^2=-1
\label{A2}
\end{eqnarray}
where, the dot denotes the differentiation with respect to the proper time.
In terms of the Killing vectors associated with the symmetries, the conserved quantities in such a spacetime can be expressed as, 
\begin{eqnarray}
-E=\textsl{g}_{tt}\dot{t}+\textsl{g}_{t\phi}\dot{\phi} \nonumber , \\ 
L=\textsl{g}_{t\phi}\dot{t}+\textsl{g}_{\phi\phi}\dot{\phi} \label{A3}
\end{eqnarray}
where, $E$ and $L$ are the energy and angular momentum per unit rest mass of the test particle moving in this spacetime. 

Let us now consider a general case, i.e., the non-geodetic motion of a test particle in the presence of a non-gravitional force. 
In such a case, the equation of motion of the test particle can be written as
\begin{equation}
    \ddot{x}^\mu+\Gamma^\mu_{\alpha\beta}\dot{x}^\alpha \dot{x}^\beta=f^\mu
    \label{geo}
\end{equation}
where, $f^\mu$ is the 4-acceleration vector. $f^\mu$ comes into play if any non-gravitational force is applied to the test particle. In the absence of such a non-grarvitational force, i.e., $f^\mu \rightarrow 0$, Eq. (\ref{geo}) reduces to the geodesic equation.
Now, if the test particle moves in a circular orbit (i.e., constant $r$) located in a constant $\theta$ plane of a general stationary and axisymmetric spacetime, the first term of Eq. (\ref{geo}) vanishes and we are left with
\begin{align}
    \Gamma^r_{tt}\dot{t}^2+2\,\Gamma^r_{\phi t}\dot{t}\,\dot{\phi}+\Gamma^r_{\phi\phi}\dot{\phi}^2=f^r \label{fr}\\
    \Gamma^\theta_{tt}\dot{t}^2+2\,\Gamma^\theta_{\phi t}\dot{t}\,\dot{\phi}+\Gamma^\theta_{\phi\phi}\dot{\phi}^2=f^\theta \label{ftheta}
\end{align}
where, $f^r$ and $f^{\theta}$ are the $r$ and $\theta$ components of the acceleration appearing due to the non-gravitational force acting on the test particle. $\Omega=\dot{\phi}/\dot{t}$ signifies the orbital angular velocity of the test particle, and, in such a case, $\dot{t}$ can be obtained from Eq. (\ref{A2}) for a constant $r$ and $\theta$. After writing the Christoffel symbols in terms of the metric components, replacing $f^r$ with $\xi$ and considering $\dot{\phi} = \Omega \dot{t}$, we obtain 
\begin{equation}
    \Omega=\frac{-\textsl{g}_{\phi t,r}+2\xi \textsl{g}_{rr}\textsl{g}_{\phi t}\pm \sqrt{\left( \textsl{g}_{\phi t,r}- 2\xi\textsl{g}_{rr}~\textsl{g}_{\phi t}\right){}^2- \left( \textsl{g}_{\phi \phi ,r}-2\xi\textsl{g}_{rr} ~\textsl{g}_{\phi \phi }\right) \left( \textsl{g}_{tt,r}-2\xi\textsl{g}_{rr} ~\textsl{g}_{tt}\right)}}{\textsl{g}_{\phi \phi ,r}-2\xi \textsl{g}_{rr}~ \textsl{g}_{\phi \phi}}
    \label{geno}
\end{equation}
by solving Eq. \eqref{fr} for $\Omega$.
The `$+$' and `$-$' signs stand for the prograde and retrograde orbits, respectively. 
Now, by substituting the expression of $\dot{t}$ (as mentioned above) in Eq. \eqref{A3}, and using the definition of $\Omega\,(=\dot{\phi}/\dot{t})$, we obtain:
\begin{eqnarray}
 E=\frac{-\textsl{g}_{tt}-\Omega \textsl{g}_{t\phi}}{\sqrt{-\textsl{g}_{tt}-2\Omega \textsl{g}_{t\phi}-\Omega^2\textsl{g}_{\phi\phi}}} \nonumber , \\ 
 L=\frac{\textsl{g}_{t\phi}+\Omega \textsl{g}_{\phi\phi}}{\sqrt{-\textsl{g}_{tt}-2\Omega \textsl{g}_{t\phi}-\Omega^2\textsl{g}_{\phi\phi}}} \label{EandL}.
\end{eqnarray}

If we consider the small perturbation,
 $r(t)=r+\delta r(t)$ \cite{staykov_orbital_2015, GMmonopoleCC2018} 
of a stable circular orbit of radius $r$ in a constant $\theta$ plane, one obtains the radial epicyclic frequency ($\nu_r$) as,
\begin{equation}\label{A5}
    \nu_r^2
   =\f{(\textsl{g}_{tt}+\Omega\textsl{g}_{t\phi})^2}{2 (2\pi)^2~\textsl{g}_{rr}}\left[\p_r^2\left({\textsl{g}_{\phi\phi}}/{Y}\right) +2l~\p_r^2\left({\textsl{g}_{t\phi}}/{Y}\right)+l^2~\p_r^2\left({\textsl{g}_{tt}}/{Y}\right) \right]\Big|_{(r\rightarrow {\rm constant},~\theta \rightarrow {\rm constant})}
\end{equation}
where, 
\begin{equation}
l=\frac{L}{E}=-\frac{\textsl{g}_{t\phi}+\Omega \textsl{g}_{\phi\phi}}{\textsl{g}_{tt}+\Omega \textsl{g}_{t\phi}}
\end{equation}
is the proper angular momentum and $Y=\textsl{g}_{tt}\textsl{g}_{\phi\phi}-\textsl{g}_{t\phi}^2$. 
By setting $\nu_r^2=0$ and solving it for $r$, one can obtain the radius of the innermost stable circular orbit (ISCO) as a function of $\theta$, which signifies the ISCO radius in that particular $\theta$ plane, and thereby we call it as `general ISCO'. $\nu_r^2 < 0$ implies the radial instabilities, i.e., these orbits are not stable. On the other hand, the circular orbits with $\nu_r^2 \geq 0$ are stable \cite{staykov_orbital_2015}.

\subsection{Application to the slowly-spinning Kerr BH inside a host star}

Applying the above-mentioned formalism, here we derive the expression for the angular momentum of a test particle which orbits a slowly-spinning Kerr BH inside a host star. In such a scenario, a non-gravitational force is applied on the test particle, which arises due to the pressure gradient terms of the host's matter. 
In this particular case, the pressure gradient components along $r$ and $\theta$ are represented by $f^r \sim -\rho^{-1} \partial P/\partial r$ (i..e., $\xi$) and $f^{\theta} \sim -\rho^{-1} \partial P/r\partial \theta$, where $P$ and $\rho$ are the pressure and density of the host star. This indicates that $-\rho^{-1} \partial P/r\partial \theta$ balances the acceleration terms of the left hand side in Eq. (\ref{ftheta}), and this, in fact, supports the circular orbits in the off-equatorial planes.

Here, we assume $\xi$ remains constant in the core of the host star for simplicity. Now, substituting the explicit forms of metric components for a slowly-spinning Kerr BH of mass $M$ and dimensionless Kerr parameter $a$, we obtain the angular velocity of the test particle in the prograde orbits as 
\begin{eqnarray}
  \Omega_{\rm K} = \sqrt{\f{\left(-\xi+\f{M}{r^2}\right)(r-2M)}{r(r-2M-\xi r^2)}}~\text{cosec}~\theta -\frac{aM^2}{r^3} +\mathcal{O}(a^2)+\mathcal{O}\left(\xi\,a\right)
  \label{ok}
\end{eqnarray}
in the linear order of $a$ by using Eq. (\ref{geno}). Substituting the expression of $\Omega_{\rm K}$ in Eq. (\ref{EandL}), we obtain the
specific angular momentum $L$ in the linear order of $a$ as
\begin{equation}
L_{\rm K}=r^2\sqrt{\f{\left(-\xi+\f{M}{r^2}\right)}{(r-3M)}} \sin \theta-\frac{3\, a \,M^2 \sin ^2\theta \, (r-2 M)}{\sqrt{r}\; (r-3 M)^{3/2}}+\mathcal{O}\left(a^2\right)+\mathcal{O}\left(\xi\,a\right).
\label{lk}
\end{equation}
Now, calculating $\nu_r^2$, and setting it to zero, one can obtain the general ISCO (gISCO) equation as
\begin{eqnarray}
   r\left[12 M^3 - 3 \xi r^4 - 4 M^2 r (2 + 3 \xi r) + M r^2 (1 + 14 \xi r)\right]+ 8 a M^2 \sqrt{Mr} ~ (r - 2M)\sin\th=0,
   \label{ISCO}
\end{eqnarray}
the solution of which is written as
\begin{equation}
    r_{\text{gISCO}}=R_{\rm S} (M, \xi)
    -4\sqrt{\frac{2}{3}}~aM\sin\theta +\mathcal{O}(a^2)+\mathcal{O}\left(\xi\,a\right)
    \label{gisco}
\end{equation}
where, $R_{\rm S} (M, \xi)$ is the exact form of gISCO radius for the Schwarzschild spacetime for a nonzero $\xi$. The exact expression of  $R_{\rm S}$ is very big, and that can be provided upon request.

Substituting the expression of $r_{\text{gISCO}}$ in Eq. \eqref{lk}, we obtain the expression of angular momentum at $r_{\text{gISCO}}$, which is denoted by
\begin{equation}\label{lgisco}
    \ell_{\text{gISCO}} \approx \sin\theta \left[R_{\rm S}^2 \sqrt{\f{\left(-\xi+\f{M}{R_{\rm S}^2}\right)}{(R_{\rm S}-3M)}} 
    -\frac{2\sqrt{2}}{3}~ aM\sin \theta \right]+\mathcal{O}\left(a^2\right)+\mathcal{O}\left(\xi\,a\right).
\end{equation}
Note that for $\theta=\pi/2$, Eq. \eqref{lgisco} reduces to Eq. (5.4.9) of \cite{novikov_astrophysics_1973} in the linear order in $a$ with $\xi \rightarrow 0$. In a similar manner, all the other related expressions (e.g., \ref{ok}-\ref{lgisco}) also reduce to the respective equations for the slowly-spinning Kerr BH with $\theta=\pi/2$ and  $\xi\rightarrow 0$. We can neglect the second term in the square bracket of Eq. \eqref{lgisco} for the case under consideration in this paper, since the Kerr parameter is very small for an EBH. In such a case, Eq. \eqref{lgisco} with  $a \rightarrow 0$ gives the exact expression of angular momentum at $R_{\rm S}$  in the Schwarzschild spacetime for a nonzero $\xi$.

\subsection{Possible numerical value of $\xi$ close to an EBH under consideration}

It is important to note here that the accreting matter around the EBH has already dynamically adjusted to the gravitational field of the EBH, as seen from Eqs. (\ref{fr}) and (\ref{ftheta}). 
In such a case, the pressure gradient should be determined by the balance with the centrifugal and gravitational accelerations. A close observation reveals that, if we solve (approximately) Eq. (\ref{ftheta}) for $f^\theta$ considering a Schwarzschild EBH, we should get it in the order of $|f^\theta | \sim M^{-1}$. 
In addition, a simple mathematical calculation by using Eq. (\ref{ISCO}) with $a \rightarrow 0$ reveals that $R_{\rm S}$ should be within $3.54M$ and $6M$ for $\infty > |\xi | > 0$.  However, as $|\xi |$ should also be in the order of $\sim M^{-1}$ (not arbitrarily higher order), $R_{\rm S}$ should be greater than $3.54M$. 
Considering the above point, we numerically find that $R_{\rm S} \sim 3.9M$ satisfies the force balance equations with numerical value of the pressure gradients around $\sim -0.1M^{-1}$. This also indicates that the above-mentioned pressure gradient values are sufficient to hold the off-equatorial orbits. 
Here we chose the value of pressure gradient ($\xi$) such that the stalled accretion as proposed here is realized. This means that the delicate balance of gravitational, centrifugal and pressure gradient accelerations is attained perfectly off the axis near $r_{\rm gISCO}$. 
Whether this value of $\xi$ is a reliable estimate for a real physical scenario is to be tested in a more detailed and realistic model of a future work.

The inward shifting of $R_S$ from the usual $6M$ for geodesic circular orbits is due to the presence of pressure-gradient forces. One can confirm this from the position of minimum of proper angular momentum $(l)$ at exactly the same $R_{\rm S}$ found by the solution of gISCO equation. 
Due to the pressure gradient forces, the radial balance equations are modified and this necessitates an increase in the centrifugal force to restore radial balance. This in essence means an increase in angular momentum. One also can observe that the pressure has a maximum at $r_{\rm gISCO}$ and hence above this, pressure gradient becomes a stabilizing force allowing for circular orbits to exist. Hence an inward shift of gISCO from the usual ISCO at $6M$ is observed in this case.
Similar cases of inwardly shifting ISCO are found in models of thick accretion disks where pressure gradients play a significant role \cite{Watarai2003, Abramowicz_1978, Paczynski1982}.

Now, considering $\xi \sim -0.1M^{-1}$ with $R_{\rm S} \sim 3.9M$, Eq. (\ref{lgisco}) reduces to

\begin{equation}\label{lgf}
\ell_{\text{gISCO}}\big|_{\xi \sim -0.1M^{-1}} \approx 6 M\sin\theta .
\end{equation}

The takeaway message from this section is that the pressure gradient terms of the particular values shown above make it possible for the circular orbits to exist in the off-equatorial plane with $\ell_{\text{gISCO}} \sim 6 M\sin\theta$. 

\bibliography{MyLibrary}
\bibliographystyle{apsrev4-2}
\end{document}